\pdfoutput=1
\documentclass{article}
\usepackage[T1]{fontenc}   % without it Tectonic/XeTeX falls back to Latin Modern instead of Times
\usepackage{spconf,amsmath,graphicx}   % official ICASSP 2027 kit (cmsworkshops.com)
\usepackage{booktabs,url,microtype}
\usepackage[hidelinks,pdftitle={Measuring the Verifier's Floor: How Much Apparent Rare-Word Text-to-Speech Error Is ASR Error?},pdfauthor={Kirill Borodin, Vasilii Kudryavtsev, Nikita Vasiliev, Grach Mkrtchian}]{hyperref}
\newcommand{\gFuncj}{0.63}
\newcommand{\gFadj}{0.22}

\newcommand{\gFfloorErr}{3.9}
\newcommand{\gFnonfloorErr}{0.2}

\newcommand{\gFrescuedN}{16}
\newcommand{\gFadjudN}{30}

\newcommand{\gFnerr}{23}

\newcommand{\gFratioCI}{$2.8$ [$2.1$, $4.1$]}
\newcommand{\gFuncSel}{0.71}
\newcommand{\gFadjSel}{0.38}

\newcommand{\gFratioSelCI}{$1.9$ [$1.5$, $2.4$]}
\newcommand{\gFflBoth}{30}
\newcommand{\gFnfBoth}{99}
\newcommand{\gFmaskj}{0.30}

\newcommand{\gFcov}{96}
\newcommand{\gFuncOne}{1.51}
\newcommand{\gFadjOne}{0.63}

\newcommand{\gFrDiffStd}{$+0.063$ [$-0.010$, $+0.145$]}

\newcommand{\gFtUncj}{$+0.14$$^*$}
\newcommand{\gFtUncjZ}{$+0.23$$^*$}
\newcommand{\gFtEqj}{$+0.04$}
\newcommand{\gFtEqjZ}{$+0.11$$^*$}
\newcommand{\gFtAdj}{$+0.08$}
\newcommand{\gFtAdjZ}{$+0.15$$^*$}
\newcommand{\gFrDiff}{$+0.037$ [$-0.034$, $+0.108$]}

\newcommand{\gKuncj}{0.72}
\newcommand{\gKadj}{0.22}

\newcommand{\gKfloorErr}{4.9}

\newcommand{\gKrescuedN}{23}
\newcommand{\gKadjudN}{36}

\newcommand{\gKnerr}{23}

\newcommand{\gKratioCI}{$3.2$ [$2.4$, $4.8$]}
\newcommand{\gKuncSel}{0.80}
\newcommand{\gKadjSel}{0.32}

\newcommand{\gKratioSelCI}{$2.5$ [$2.0$, $3.4$]}

\newcommand{\gKmaskj}{0.36}

\newcommand{\gKuncOne}{2.26}
\newcommand{\gKadjOne}{0.43}

\newcommand{\gKrDiffStd}{$+0.032$ [$-0.045$, $+0.107$]}

\newcommand{\gKtUncj}{$+0.16$$^*$}
\newcommand{\gKtUncjZ}{$+0.28$$^*$}
\newcommand{\gKtEqj}{$+0.06$}
\newcommand{\gKtEqjZ}{$+0.13$$^*$}
\newcommand{\gKtAdj}{$+0.13$$^*$}
\newcommand{\gKtAdjZ}{$+0.18$$^*$}
\newcommand{\gKrDiff}{$+0.068$ [$-0.017$, $+0.154$]}

\newcommand{\fluncj}{1.52}
\newcommand{\fladj}{0.46}

\newcommand{\flnerr}{18}

\newcommand{\flratioCI}{$3.3$ [$2.3$, $5.6$]}
\newcommand{\fluncSel}{1.62}
\newcommand{\fladjSel}{0.76}

\newcommand{\flratioSelCI}{$2.1$ [$1.6$, $3.2$]}

\newcommand{\flmaskj}{0.66}

\newcommand{\flcov}{94}
\newcommand{\fluncOne}{2.56}
\newcommand{\fladjOne}{1.11}

\newcommand{\lsVRhumanC}{32.9}
\newcommand{\lsComHumanC}{1.5}
\newcommand{\lsVRhumanA}{31.8}
\newcommand{\lsVRhumanB}{30.3}
\newcommand{\lsVRunjudged}{21.6}

\newcommand{\lsVRadj}{3.4}

\newcommand{\lsVRerr}{10}

\newcommand{\lsComAdjN}{0 of 7,889}
\newcommand{\lsVRadjN}{10 of 298}
\newcommand{\lsVRallN}{3 of 215}

\newcommand{\lsVRoneErr}{4}
\newcommand{\lsVRoneN}{32}

\newcommand{\flVRhumanC}{66.7}
\newcommand{\flComHumanC}{3.2}

\newcommand{\flVRunjudged}{41.7}

\newcommand{\lsKVRadj}{4.7}

\newcommand{\lsKVRerr}{14}

\newcommand{\lsKComAdjN}{3 of 7,889}
\newcommand{\lsKVRadjN}{14 of 298}

\newcommand{\lsKVRoneErr}{5}

\newcommand{\accLeMin}{66}
\newcommand{\accLeMax}{75}
\newcommand{\neverFloorMin}{69}
\newcommand{\neverFloorMax}{80}
\newcommand{\binSizes}{124, 124, 111 and 141}
\newcommand{\binLenMin}{20.0}
\newcommand{\binLenMax}{21.5}

\newcommand{\gFresN}{23}
\newcommand{\gFresM}{15}
\newcommand{\gFresS}{8}

\newcommand{\gFadjMisp}{0.15}
\newcommand{\gFVRadjMisp}{1.7}
\newcommand{\gKresN}{23}
\newcommand{\gKresM}{8}
\newcommand{\gKresS}{15}

\newcommand{\gKadjMisp}{0.08}
\newcommand{\gKVRadjMisp}{1.7}
\newcommand{\flresN}{18}
\newcommand{\flresM}{12}
\newcommand{\flresS}{6}

\newcommand{\fladjMisp}{0.30}
\newcommand{\flVRadjMisp}{5.7}
\newcommand{\proxyR}{0.60}
\newcommand{\freqLSrows}{very rare ($z{<}2.5$) & 380 & 21.6 & 32.9 & 21.8 / 25.0 & 10.1 / 13.4 & $3.4$ [$1.4$, $5.4$] & $4.7$ [$2.5$, $7.2$] \\
rare ($2.5$--$3.5$) & 783 & 4.3 & 8.9 & 4.3 / 2.6 & 2.8 / 1.1 & $1.5$ [$0.7$, $2.4$] & $0.4$ [$0.0$, $0.9$] \\
mid ($3.5$--$4.5$) & 1319 & 1.4 & 4.7 & 0.9 / 1.1 & 0.5 / 0.8 & $0.2$ [$0.0$, $0.4$] & $0.2$ [$0.0$, $0.4$] \\
common ($z{\ge}4.5$) & 7930 & 0.5 & 1.5 & 0.2 / 0.3 & 0.1 / 0.2 & $0.0$ [$0$, $0.05$]$^\dagger$ & $0.0$ [$0.0$, $0.1$] \\}
\newcommand{\freqFLrows}{very rare ($z{<}2.5$) & 60 & 41.7 & 66.7 & 50.0 & 31.4 & $11.4$ [$2.4$, $24.1$] & --- \\
rare ($2.5$--$3.5$) & 185 & 11.9 & 27.0 & 15.1 & 10.4 & $1.8$ [$0.0$, $4.6$] & --- \\
mid ($3.5$--$4.5$) & 619 & 3.2 & 7.8 & 3.2 & 2.3 & $1.0$ [$0.2$, $2.1$] & --- \\
common ($z{\ge}4.5$) & 3096 & 1.3 & 3.2 & 0.5 & 0.4 & $0.1$ [$0.0$, $0.3$] & --- \\}
\newcommand{\maskrows}{LS, F5-TTS & $3.9$ [$1.5$, $6.8$] & $0.2$ [$0.1$, $0.3$] & $53$ [$35$, $71$] \\
LS, Kokoro & $4.9$ [$2.2$, $8.1$] & $0.2$ [$0.1$, $0.3$] & $64$ [$48$, $79$] \\
FLEURS, F5-TTS & $6.0$ [$2.3$, $10.4$] & $0.3$ [$0.1$, $0.6$] & $56$ [$35$, $79$] \\}
\newcommand{\gFrecov}{54}
\newcommand{\flrecov}{55}
\newcommand{\nLS}{500}
\newcommand{\meanWords}{20.8}
\newcommand{\minWords}{15}
\newcommand{\maxWords}{30}
\newcommand{\densMax}{41}
\newcommand{\firstN}{300}
\newcommand{\extN}{200}

\newcommand{\firstAbove}{37}

\newcommand{\extMin}{18}
\newcommand{\extMax}{41}
\newcommand{\flRecordings}{200}
\newcommand{\flSentences}{161}
\newcommand{\flRepeated}{39}
\title{Measuring the Verifier's Floor: How Much Apparent Rare-Word\\
       Text-to-Speech Error Is {ASR} Error?}
\name{Kirill Borodin$^{1}$, Vasilii Kudryavtsev$^{1}$, Nikita Vasiliev$^{1}$, Grach Mkrtchian$^{2}$}
\address{$^{1}$BitmanagerAI, Dubai, UAE\qquad
         $^{2}$lab260, Yerevan, Armenia\\
         \small\texttt{kborodin.research@gmail.com}}

\usepackage{eso-pic}
\AddToShipoutPictureFG*{%
  \put(0,34){\makebox[\paperwidth]{\parbox{6.4in}{\centering\footnotesize
    This work has been submitted to the IEEE ICASSP for possible publication.
    Copyright may be transferred without notice, after which this version
    may no longer be accessible.}}}}

\begin{document}
\maketitle

\begin{abstract}
Text-to-speech (TTS) is usually evaluated by transcribing its output with an
automatic speech recognition (ASR) system, so the word error rate also counts
recognizer errors. We measure these errors on human recordings of the test
sentences. On LibriSpeech and FLEURS, the evaluation recognizer misses
33--67\% of very rare words even in human speech, and all three recognizers miss
22--42\%. Checking every word with two more recognizer families lowers the apparent
TTS error on words that at least one recognizer hears two- to three-fold (very rare
LibriSpeech words: from 10--13\% to 3--5\%), yet very rare words remain the hardest.
Ignoring words the recognizer misses in human speech is biased: the other
recognizers find TTS errors far more often on these words (4--6\% vs.\
0.2--0.3\%) and overturn about half of the errors this mask keeps. As half of the
remaining adjudicated errors may be spelling variants, a listening test is needed.
\end{abstract}

\section{Introduction}
\label{sec:intro}
The standard test of whether a text-to-speech (TTS) system says the right words is
the \emph{ASR round-trip}: the synthetic speech is transcribed by a speech
recognizer, and the word error rate (WER) is computed against the input
text~\cite{seedtts2024,f5tts2024}. Recent systems also use the recognizer at
inference: they generate several candidates and keep the one it transcribes
best~\cite{llasa2025,selfverify2026,familyalign2026}.

This test measures two systems at once. If the recognizer cannot transcribe a word
even when a person says it, an error on this word tells us little about the TTS
system. Such words are mostly rare words, names and foreign words, which is exactly
where TTS errors are reported to concentrate.

A recent study~\cite{selfverify2026} calls rare words
``a genuine limit rather than a sampling problem'', but notes that it ``did not
measure its false-positive floor on ground-truth audio''. We measure this floor. We use corpora in which every test sentence
has its original human recording, and we run the same recognizers on human and on
synthetic speech. The error on human speech is a diagnostic baseline, not a strict
bound, because synthetic speech can be easier or harder to recognize.

We compare three estimates of TTS word error: the standard round-trip, a
\emph{single-recognizer mask} that ignores words the recognizer misses in human
speech, and a \emph{three-recognizer adjudication} in which two more recognizers
check every word. Our main findings are:
\begin{enumerate}\itemsep0pt
\item The recognizer error depends strongly on word frequency. In human LibriSpeech
      recordings it misses \lsComHumanC\% of common but \lsVRhumanC\% of very rare
      words.
\item Adjudication lowers the apparent TTS error two- to three-fold. Very rare words
      remain the hardest, though about half of the remaining errors may be spelling
      variants.
\item The single-recognizer mask is biased. It removes the words on which the other
      recognizers find TTS errors most often, and they overturn half of those it keeps.
\end{enumerate}

\section{Related work}
\label{sec:related}
ASR transcripts are widely used instead of listener transcripts to test TTS
intelligibility~\cite{taylor2021}, although ASR WER agrees only loosely with
listeners~\cite{responsible2025}. Yu and Kang~\cite{familyalign2026} show that
best-of-$N$ results depend on the recognizer family used as judge. Luo and
Wan~\cite{maskedchinese2026} report the opposite error to ours: a recognizer can hide
real TTS mistakes by writing what the text should say.

The closest work is Appendix~C of~\cite{responsible2025v2} (not in the ICML
version~\cite{responsible2025}). It compares ASR WER with listener transcripts of
synthetic and human speech for one TTS system, at corpus level. We add the
dependence on word frequency, word-level checks with three recognizer families, and
their effect on rare-word results such as~\cite{selfverify2026}, whose system we do
not re-run.

\section{Method}
\label{sec:method}

\subsection{Word decisions}
For each test sentence we have its text, the original human recording and $N$
synthetic versions. Three recognizers, A, B and C, transcribe all of them. C is the
evaluator (the verifier) whose round-trip we study; A and B are only used to check C.

Texts are normalized with the Whisper English text normalizer and aligned. A
reference word is \emph{recovered} (the recognizer \emph{hears} it) if it matches
exactly, or if up to three words differ from the transcript only in spaces
(\emph{any thing} vs.\ \emph{anything}). We ignore insertions, so our rates are word
miss rates, not WER.

For recognizer $X$, $h_X(w)$ is true if $X$ hears word $w$ in the human recording,
and $s_X(w)$ is true if $X$ hears $w$ in at least one of the $N$ synthetic versions.
Our rates are therefore any-of-$N$ rates, which fall as $N$ grows. We also report
$N{=}1$ and the one candidate that a best-of-$N$ system would keep.

\subsection{Three estimates of TTS error}
Word $w$ is a TTS error under
\emph{(1)~the standard round-trip} if $\neg s_C(w)$, and under
\emph{(2)~the single-recognizer mask} if $h_C(w)\wedge\neg s_C(w)$, that is, words
that C misses in human speech are ignored. Under
\emph{(3)~three-recognizer adjudication},
\begin{equation}
\begin{split}
\mathrm{err}(w)={}&\big[h_A(w)\lor h_B(w)\lor h_C(w)\big]\\
                  &\wedge\neg\big[s_A(w)\lor s_B(w)\lor s_C(w)\big].
\end{split}
\label{eq:adj}
\end{equation}
A word is \emph{judged} if at least one recognizer hears it in the human recording.
Under (3), a judged word is an error only if no recognizer hears it in any synthetic
version. We report all three estimates per judged word.

The mask can fail in two ways. The ignored words may also be TTS errors, and an
error it keeps may still be a mistake of C. Adjudication reduces both problems but
does not remove them. It leaves out words that no recognizer hears, and a recognizer
may guess a word from context~\cite{maskedchinese2026}; both lower the error.
Spelling variants and shared recognizer failures raise it.

\begin{table}[t]\centering\small
\caption{Four words from the F5-TTS LibriSpeech set: recognizers that hear the word in
the human recording, the most frequent transcript of A / B / C over the 8 synthetic
versions (=: correct), and the decision of estimates (1)--(3).}
\label{tab:example}
\setlength{\tabcolsep}{2pt}
\begin{tabular}{@{}lllccc@{}}
\toprule
word & human & TTS output (A / B / C) & (1) & (2) & (3) \\
\midrule
Phoebe & A, B, C & fuby / pube / fuby   & error & error   & error \\
Tessa  & A, B, C & = / = / tess         & error & error   & correct \\
stile  & A, B    & = / = / style        & error & ignored & correct \\
Veolan & A       & violan / violan / tullyviolen & error & ignored & error \\
\bottomrule
\end{tabular}
\end{table}

Table~\ref{tab:example} shows four real words. All recognizers mishear
\emph{Phoebe} in every synthetic version, so all estimates count an error. For
\emph{Tessa} and \emph{stile} only C fails in the synthetic speech, so adjudication
counts no error.
\emph{Veolan} is heard only by A in the human recording and by no recognizer in the
synthetic speech: the mask ignores it, but adjudication counts it.

\subsection{Word frequency}
We measure word frequency with the wordfreq library~\cite{wordfreq} on the Zipf
scale, where $z{=}3$ means once per million words. \emph{Very rare} words have $z{<}2.5$ and
\emph{common} words $z{\ge}4.5$. Tokens with digits are not assigned to a band.

The LibriSpeech sentences were selected with a simpler corpus measure: a word is rare if it
occurs at most twice in about 11k LibriSpeech transcripts. This measure is also the $x$-axis
of Fig.~\ref{fig:grad}. Per sentence, it correlates with the share of words below
$z{=}3$ at $r{=}\proxyR$, and we repeat the trend analysis with this Zipf share.

\section{Experimental setup}
\label{sec:setup}
\begin{figure*}[t]
\centering
\includegraphics[width=\textwidth]{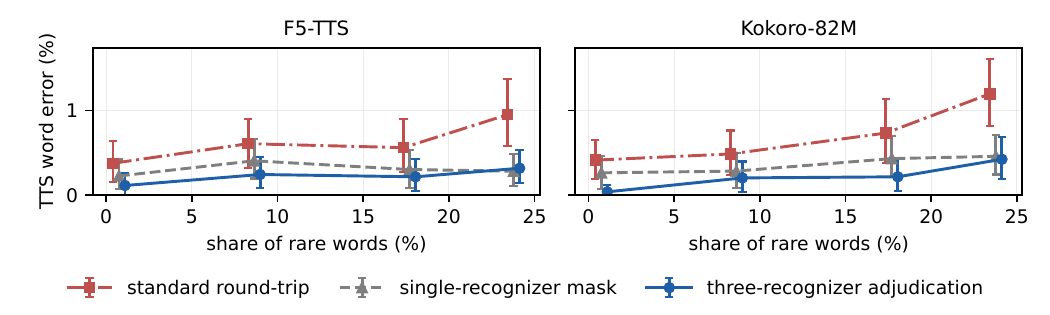}  % drawn at 7 in: glyphs stay >= 9 pt
\caption{TTS word error per judged word (any-of-$N$) against the share of rare words
(corpus measure) in the \nLS{} LibriSpeech sentences, split into four bins at the
quartiles (\binSizes{} sentences). Points are bin means; bars are 95\% CIs.}
\label{fig:grad}
\end{figure*}

\textbf{Data.} The main set has \nLS{} LibriSpeech~\cite{librispeech2015} test and
dev sentences of \minWords--\maxWords{} words (mean \meanWords). The share of rare
words varies from $0$ to \densMax\%, while the mean length stays between
\binLenMin{} and \binLenMax{} words across the bins of Fig.~\ref{fig:grad}.

The first \firstN{} sentences had only \firstAbove{} with more than $15\%$ rare words
and gave unclear trends. We then added the \extN{} remaining sentences with the
highest shares (\extMin--\extMax\%). This was decided after seeing the first
results, so the trend analysis is exploratory.

We repeat the measurement on the first \flRecordings{} FLEURS en\_us~\cite{fleurs}
test recordings with $12$--$30$ words and $3$--$16$\,s of audio. They cover
\flSentences{} sentences (\flRepeated{} recorded twice) with more names and numbers.

\textbf{Systems.} F5-TTS~\cite{f5tts2024} generates $N{=}8$ versions with random
seeds ($N{=}4$ on FLEURS) and clones the voices of five LibriSpeech test-clean speakers.
Kokoro-82M~\cite{kokoro2025} is deterministic, so its $N{=}4$ versions use four
voices. Since $N$ and the kind of variation differ, we do not compare the two systems.

The recognizers are A:~Whisper large-v3-turbo~\cite{whisper2023}, B:~wav2vec\,2.0
large~\cite{wav2vec2020} (CTC, no language model) and C:~OWSM v3.1~\cite{owsm2024},
all with greedy decoding.

\textbf{Statistics.} Words in a sentence are not independent, so 95\% confidence
intervals (CIs) come from a percentile bootstrap over sentences (4,000 resamples).
Repeated FLEURS sentences are resampled together. A cell with no errors gets the
95\% Clopper--Pearson upper limit, which assumes independent words.

Trends are Pearson correlations over sentences between the share of rare words and
the error per judged word. No parameters were fitted to these data, but the word
matching rule was revised after inspecting alignments from these sets.

\section{Results}
\label{sec:results}

\begin{table*}[t]\centering\small
\caption{ASR error on human speech and TTS error by word frequency (Zipf band $z$;
tokens with digits excluded). \emph{Unjudged}: no recognizer hears the word in the
human recording. \emph{Standard}: C misses the word in all $N$ synthetic versions
(LibriSpeech: F5-TTS / Kokoro). \emph{Adjudicated}: Eq.~(\ref{eq:adj}), with 95\% CIs;
$^\dagger$no errors, 95\% Clopper--Pearson upper limit. All values in \%.}
\label{tab:freq}
\setlength{\tabcolsep}{4pt}
\begin{tabular}{@{}lrcccccc@{}}
\toprule
 & & \multicolumn{2}{c}{ASR on human speech} & \multicolumn{2}{c}{standard TTS error} & \multicolumn{2}{c}{adjudicated TTS error} \\
\cmidrule(lr){3-4}\cmidrule(lr){5-6}\cmidrule(lr){7-8}
band & words & unjudged & C misses & all words & judged words & F5-TTS & Kokoro \\
\midrule
\multicolumn{8}{@{}l}{\emph{LibriSpeech, \nLS{} sentences}}\\
\freqLSrows
\midrule
\multicolumn{8}{@{}l}{\emph{FLEURS en\_us, F5-TTS, \flRecordings{} recordings of \flSentences{} sentences}}\\
\freqFLrows
\bottomrule
\end{tabular}
\end{table*}

\subsection{ASR errors on human speech}
Table~\ref{tab:freq} shows that the recognizer error on human speech grows quickly
as words get rarer. C misses \lsComHumanC\% of common LibriSpeech words but
\lsVRhumanC\% of very rare ones (\flComHumanC\% and \flVRhumanC\% on FLEURS). On LibriSpeech, A and B also miss \lsVRhumanA\% and \lsVRhumanB\% of them.

For \gFrecov\% of the words that C misses (\flrecov\% on FLEURS), A or B does
transcribe them from the same recording. This points to a problem of C rather than
of the audio, although A or B may also guess a word from context. The rest may be
unclear audio, dialect spellings that all recognizers normalize (\emph{hisself}
$\to$ \emph{himself}) or shared blind spots.

All three recognizers miss \lsVRunjudged\% of the very rare LibriSpeech words
(\flVRunjudged\% on FLEURS), so these words cannot be judged at all.

\begin{table}[t]\centering\small
\caption{TTS word error per judged word (\%) and its correlation $r$ with the share
of rare words per sentence (corpus measure/Zipf share; $^*$: the 95\% CI excludes
zero). \emph{Kept}: the version with the fewest words missed by C.}
\label{tab:est}
\setlength{\tabcolsep}{2pt}
\begin{tabular}{@{}llccc@{}}
\toprule
 & & \multicolumn{2}{c}{LibriSpeech} & FLEURS \\
\cmidrule(lr){3-4}
 & estimate & F5-TTS & Kokoro & F5-TTS \\
\midrule
any-of-$N$ & standard    & \gFuncj  & \gKuncj  & \fluncj \\
           & mask        & \gFmaskj & \gKmaskj & \flmaskj \\
           & adjudicated & \gFadj   & \gKadj   & \fladj \\
\midrule
$N{=}1$    & standard    & \gFuncOne & \gKuncOne & \fluncOne \\
           & adjudicated & \gFadjOne & \gKadjOne & \fladjOne \\
\midrule
kept       & standard    & \gFuncSel & \gKuncSel & \fluncSel \\
           & adjudicated & \gFadjSel & \gKadjSel & \fladjSel \\
\midrule
$r$        & standard    & \gFtUncj/\gFtUncjZ & \gKtUncj/\gKtUncjZ & -- \\
           & mask        & \gFtEqj/\gFtEqjZ   & \gKtEqj/\gKtEqjZ   & -- \\
           & adjudicated & \gFtAdj/\gFtAdjZ   & \gKtAdj/\gKtAdjZ   & -- \\
\bottomrule
\end{tabular}
\end{table}

\subsection{TTS error under the three estimates}
Table~\ref{tab:est} compares the estimates on the same judged words. With
any-of-$N$, adjudication lowers the error from \gFuncj\% to \gFadj\% for F5-TTS and
from \gKuncj\% to \gKadj\% for Kokoro. The ratio of standard to adjudicated error is
\gFratioCI{} and \gKratioCI{} (\flratioCI{} on FLEURS).

By definition the adjudicated error cannot exceed the standard error, so the size of
the gap is the result. Part of it comes from pooling, because an adjudicated error
must survive $3N$ transcripts instead of $N$.

The gap remains for a single candidate
and for the candidate a best-of-$N$ system would keep. For the latter the ratio is
\gFratioSelCI{} and \gKratioSelCI{} (\flratioSelCI{} on FLEURS); this is
conservative, as the selection also removes some of C's own errors.

Rare words remain the hardest. On very rare LibriSpeech words, the adjudicated error
is \lsVRadj\% for F5-TTS (\lsVRadjN) and \lsKVRadj\% for Kokoro (\lsKVRadjN), against
\lsComAdjN{} and \lsKComAdjN{} common words (Table~\ref{tab:freq}).

Part of this evidence is weak. \lsVRoneErr{} of the \lsVRerr{} and \lsKVRoneErr{} of
the \lsKVRerr{} very rare errors fall on the \lsVRoneN{} words that only one
recognizer hears in the human recording. On the words that all three hear, each
system has errors on \lsVRallN{} words, mostly spelling variants
(Sec.~\ref{sec:residual}).

The error also rises with the share of rare words in a sentence
(Fig.~\ref{fig:grad}, Table~\ref{tab:est}). The adjudicated trend is significant for
Kokoro with the corpus measure and for both systems with the Zipf share. With the corpus measure, it does not differ significantly from the standard trend (\gFrDiffStd{} and \gKrDiffStd).

\subsection{What the remaining errors are}
\label{sec:residual}
There are \gFresN{}, \gKresN{} and \flresN{} adjudicated errors in the three sets,
mostly names, archaic or foreign words and, on FLEURS, abbreviations.

In \gFresM{}, \gKresM{} and \flresM{} cases the recognizer transcripts point to a different
pronunciation, for example \emph{Phoebe} as ``fuby'' or \emph{BCE} as ``beasts''.
The other \gFresS{}, \gKresS{} and \flresS{} are spelling variants or homophones
(\emph{Bessy}/\emph{Bessie}, \emph{buoy}/\emph{boy}) or text artifacts.
No transcript-based metric can tell these apart from mispronunciations.

Counting only the first group, the TTS error is \gFVRadjMisp\%, \gKVRadjMisp\% and
\flVRadjMisp\% on very rare words and \gFadjMisp\%, \gKadjMisp\% and \fladjMisp\%
overall, so rare words are still the hardest.

\begin{table}[t]\centering\small
\caption{Test of the single-recognizer mask with A and B as judges, counting only
words that A or B hears in human speech. Left: TTS error on words that C misses or
hears there. Right: share of the mask's kept errors that A or B hears in a synthetic
version. Values in \% with 95\% CIs.}
\label{tab:mask}
\setlength{\tabcolsep}{2pt}
\begin{tabular}{@{}lccc@{}}
\toprule
 & \multicolumn{2}{c}{TTS error on words C\ldots} & kept errors \\
\cmidrule(lr){2-3}
 & misses on human & hears on human & A/B hear \\
\midrule
\maskrows
\bottomrule
\end{tabular}
\end{table}

\subsection{The single-recognizer mask is biased}
Table~\ref{tab:mask} uses A and B as judges. On the words that C misses in human
speech, they find a TTS error in \gFfloorErr\% (F5-TTS) and \gKfloorErr\% (Kokoro) of
cases, but only in \gFnonfloorErr\% of the words that C hears. The mask thus removes the words that TTS gets wrong most
often, by more than an order of magnitude.

The mask also keeps false errors. Of its kept errors on words that A or B hears in
human speech, A or B hears \gFrescuedN{} of \gFadjudN{} (F5-TTS) and \gKrescuedN{} of
\gKadjudN{} (Kokoro) in a synthetic version, so adjudication overturns about half.
Words that C misses are also hard for A and B, who both hear only \gFflBoth\% of them
in human speech (against \gFnfBoth\% of the words C hears), so part of the contrast
may be shared difficulty.

Computing the mask only over the words C hears (\flcov--\gFcov\% of words) changes
its rate by less than $0.03$ points, so the denominator does not drive Table~\ref{tab:est}.
With the corpus measure, neither the mask's trend nor the difference between the
adjudicated and the mask trend (\gFrDiff{} and \gKrDiff) is significant, so we claim
only that the mask is biased.

\section{Discussion}
\label{sec:discussion}
We first planned to adapt $N$ to each sentence and to stop when C hears all words of
a candidate. With this rule, \accLeMin--\accLeMax\% of the sentences (recordings on
FLEURS) are accepted by the second candidate. Of the sentences never accepted,
\neverFloorMin--\neverFloorMax\% contain a word that C misses in the human recording
and in all $N$ versions, so C cannot tell whether TTS or C is at fault.

Our results suggest four practices for round-trip evaluation: (i)~report the
recognizer error on the human recordings, by word frequency; (ii)~do not simply drop
the words that the recognizer misses there; (iii)~check disputed rare words with
other recognizer families or, better, with listeners; (iv)~report $N{=}1$ next to
any-of-$N$.

\section{Limitations}
\label{sec:lim}
\textbf{No listening test.} All TTS-error decisions come from recognizers, which
share blind spots. A listening test on these \gFresN{}+\gKresN{}+\flresN{} errors, a sample of removed
errors and control words has been prepared but not yet run.

\textbf{Few errors.} With \gFnerr{}, \gKnerr{} and \flnerr{} adjudicated errors, the
rare-word estimates and trends have wide intervals.

\textbf{Scope.} We use English read speech from two corpora. B and C were trained on
LibriSpeech, and C is both evaluator and judge.

\section{Conclusion}
An ASR round-trip scores the recognizer as well as the TTS system, and the
recognizer's own errors are concentrated on rare words. With three recognizers on
human and synthetic speech, the apparent TTS error on judged words is two to three
times the adjudicated error, very rare words remain the hardest, and ignoring words
one recognizer misses is a biased correction that discards true errors and keeps
false ones. A listening test is needed to confirm
these results. We will release code, sentence lists and all transcripts.

\vfill\pagebreak
% ICASSP 2027 kit: the 5th page may contain only references, funding acknowledgements
% and the Compliance with Ethical Standards statement. ICASSP 2027 SPS policies: the
% conflict-of-interest disclosure goes in the Acknowledgments, which may be on this page.
\section{Compliance with Ethical Standards}
This research study was conducted retrospectively using human subject data made
available in open access under the CC BY 4.0 license by
LibriSpeech~\cite{librispeech2015} and FLEURS~\cite{fleurs}. No new recordings were
made and no listening tests were run, so ethical approval was not required.

\section{Acknowledgments}
The authors have no relevant financial or non-financial interests to disclose.
Claude Opus 5.5 (Anthropic) was used only to polish the text of this paper. All
ideas and experiments were conceived and carried out by the authors, who take full
responsibility for the content.

\bibliographystyle{IEEEbib}
\bibliography{refs}
\end{document}